\documentclass[aps,twocolumn]{revtex4}
\usepackage{epsfig}

\DeclareMathAlphabet{\bi}{OML}{cmm}{b}{it}

\newcommand{\p}{\partial}

\newcommand{\e}{\mathrm{e}}

\begin{document}

\title{Quantum Dots in a Strong Magnetic Field. \\
Quasi-classical consideration}
\date{\today}
\author{A.~Matulis}
\email[E-mail me at: ]{amatulis@takas.lt}
\affiliation{Institute of Semiconductor Physics,
Go\v{s}tauto 11, 2600 Vilnius, Lithuania}

\begin{abstract}
The electron motion in rather strong magnetic fields (when only the
lowest Landau level is populated) is considered. In this case the
electron kinetic energy is frozen out and the electrons are guided
by slowly varied potential. Using the  adiabatic procedure and
expansion in magnetic length series the approximate description
is developed. In zero order this approximation leads to the classical
equations of motion describing the Larmor circle drift in the potential
gradient. In the second order the special quantum mechanical
description where the electron potential energy plays the role
of the total Hamiltonian is constructed. Simple examples of a
single and two electrons in the parabolic dot demonstrates that
the proposed approximate description gives the main features of the
electron system spectrum and the collective phenomena.
\end{abstract}


\maketitle

\section{Introduction}

Quantum dots, or \textit{artificial atoms}, have been a subject of
intense theoretical and experimental research over the last few
years \cite{jacak98}. The useful instrument in spectroscopy
experiments is the magnetic field applied in perpendicular to the
quantum dot plane direction which enables to trace easily the
dependence of the quantum dot properties on various parameters.
Moreover the strong magnetic field reveals the quantization effects
introducing into the electron system the favorable interplay
between confining potential and Landau levels.

Recently the main interest in quantum dots is related to the
electron-electron interaction and the collective phenomena,
such as the change of the ground state multiplicity,
the electron density reconstruction, and the
Wigner crystallization. The electron density reconstruction in the
finite electron systems was considered in \cite{wen94}. Now it is
known as Shamon-Wen edge --- some of the electron density ring around the
finite system. Under certain circumstances the ring was reported to
become unstable \cite{ren99}, and it breaks into separate lumps.
Although the possibility to obtain the symmetry braking solutions
was argued \cite{wingr99} considering them as an artifacts of the
approximate methods used, the exact calculations of the electron
correlation function \cite{maksym96} undoubtedly indicates that
the Wigner crystallization occurs at rather large electron-electron
interaction. The presented in \cite{riemann99} electron density plots
show that the strong magnetic field facilitates the electron
density edge reconstruction leading to the Wigner crystallization.

Meanwhile the minimization of the system potential presented in \cite{bedanov94}
shows that the Wigner crystallization in quantum dots can be successfully
considered by classical mechanics. The fact that the strong magnetic
field facilitates the Wigner crystallization enables to suppose that
the electron system behavior in very strong magnetic can be described
by classical or quasi-classical methods. The purpose of the present paper
is to show how such methods could be developed.
The paper is organized as follows. After the formulation of the model
in the next Section, in Sections 3 and 4 the main instrument ---
fast and slow variables are introduced. Then in Section 5 the adiabatic
procedure is discussed and the slow motion Schr\"{o}dinger equation
is considered. In Section 6 the classical equations for the limit
case of strong magnetic field are derived, and in the next two Sections
the illustrations of the simplified quantum mechanical description
are given. In Appendix A the details of the adiabatic procedure
are presented, and in Appendix B the transformation back to the
initial coordinates is discussed.

\section{Model}

We consider the Schr\"{o}dinger equation
\begin{equation}\label{sred}
  i\hbar\frac{\p}{\p t} \Psi = H_T\Psi
\end{equation}
with the Hamiltonian
\begin{equation}\label{ham}
  H_T = \frac{1}{2m}\left\{\bi{p}+\frac{e}{c}\bi{A}(\bi{r})\right\}^2
  + V(\bi{r})
\end{equation}
describing the motion of 2D electrons in the strong perpendicular
homogeneous magnetic field and slowly varying potential $V(\bi{r})$.
For the sake of simplicity the main equations will be derived
for a single electron as the generalization for the system of many
electrons is trivial. It will be presented at the end of derivation.

Choosing the symmetric gauge $\bi{A}=[\bi{B}\times\bi{r}]$ we write
down the main part of the Hamiltonian as follows:
\begin{equation}\label{zeroham}
  H_0 = \frac{1}{2m}\left\{\left(p_x - \frac{eB}{2c}y\right)^2
  + \left(p_y + \frac{eB}{2c}x\right)^2\right\}.
\end{equation}
We shall consider it as a largest one treating the remaining potential
$V(\bi{R})$ as a small perturbation.

\section{Landau levels}

As in the standard perturbation technique we have to start with the
zero order problem and solve the following stationary Schr\"{o}dinger
equation:
\begin{equation}\label{zeroorder}
  \{H_0-\varepsilon\}\psi = 0.
\end{equation}
The solution of it is known as Landau levels. The most simple way
to obtain it is to introduce the new variables
\begin{equation}\label{newvar}
  \xi = \frac{l_B}{\hbar}p_x-\frac{1}{2l_B}y, \quad
  \eta = \frac{l_B}{\hbar}p_y+\frac{1}{2l_B}x
\end{equation}
where $l_B=\sqrt{c\hbar/eB}$ is the magnetic length.
Using the new variables Hamiltonian (\ref{zeroham}) can be rewritten
as
\begin{equation}\label{zhamnew}
  H_0 = \frac{\hbar\omega_c}{2}(\xi^2+\eta^2)
\end{equation}
where $\omega_c=eB/mc$ is the cyclotron frequency, and
the new variables obey the following commutation rule:
\begin{equation}\label{comrulefast}
  [\xi,\eta] = -i.
\end{equation}

The zero order Hamiltonian reminds the Hamiltonian of the harmonic
oscillator, and it is evident that it has the equidistant discrete
spectrum which as it was already mentioned is called the Landau levels.

We shall consider the case of very strong magnetic field when the
electrons are in the lowest Landau level. Our task is to reveal how
the slowly varying additional potential $V(\bi{r})$ (as compared
with the magnetic length $l_B$) changes their behavior.

\section{Slow variables}

We shall treat the variables introduced in the previous Section
as \textit{fast} variables because they are included into the
main part of the Hamiltonian. But as we are going to solve the 2D
problem they are not sufficient to treat initial Schr\"{o}dinger
equation (\ref{sred}). We have to introduce two more variables. We
shall do that in the following way:
\begin{equation}\label{slowvar}
  X = \frac{1}{2}x - \frac{l_B^2}{\hbar}p_y, \quad
  Y = \frac{1}{2}y + \frac{l_B^2}{\hbar}p_x.
\end{equation}
We chose them in such way in order to have the most simple commutation
relations, namely,
\begin{equation}\label{komsalgl}
  [\xi,X] = [\xi,Y] = [\eta,X] = [\eta,Y] = 0,
\end{equation}
and
\begin{equation}\label{komsall}
  [Y,X] = -il_B^2.
\end{equation}
We shall consider those variables as \textit{slow} ones.

Now substituting the initial variables
\begin{equation}\label{oldvar}
  x = X + l_B\eta, \quad y = Y - l_B\xi
\end{equation}
into Hamiltonian (\ref{ham}) we arrive at the following
expression:
\begin{equation}\label{hamfin}
  H_T = \frac{\hbar\omega_c}{2}(\xi^2+\eta^2) + V(X + l_B\eta,Y - l_B\xi).
\end{equation}
So, we divided the Hamiltonian into two parts. The first largest one
describing the motion of the electron in the homogeneous magnetic
field depends on the fast variables only, while the other one ---
the slowly varying potential --- depends on both fast and slow variables.
Thus, we see that the slow and fast variables can not be separated exactly,
but the presence of the small parameter (namely, the ratio of the
magnetic length $l_B$ and the characteristic potential variation length
$l_0\sim |V/\nabla V|$)
enables us to separate them approximately by means of some adiabatic
procedure.

\section{Adiabatic procedure}
\label{adiabatproc}

Now we are going to develop some adiabatic procedure and apply it for
considering the Schr\"{o}dinger equation (\ref{sred}).  For this
purpose we shall make the following steps:
\begin{itemize}
\item{we expand the potential into $l_B$-powers:
\begin{eqnarray}\label{letpotskl}
  V &=& V(X,Y) + l_B\eta V_X(X,Y) \nonumber \\
  &-& l_B\xi V_Y(X,Y) + \cdots;
\end{eqnarray}
}
\item{divide the Hamiltonian into two parts:
\begin{eqnarray}\label{hampad}
  H &=& H_f + H_s, \\
\label{hampad2}
  H_f &=& \frac{\hbar\omega_c}{2}(\xi^2+\eta^2) \nonumber \\
  &+& l_B\eta V_X(X,Y) + l_B\xi V_Y(X,Y) + \cdots, \\
  H_s &=& V(X,Y);
\end{eqnarray}
}
\item{present the wave function as the product of its fast and slow
parts:
\begin{equation}\label{bangfunk}
  \Psi = \psi(\eta|X,Y)\Phi(X),
\end{equation}
}
\item{and use the following equation for the fast wave function
part:
\begin{equation}\label{greitlg}
  \{H_f-E(X,Y)\}\psi(\eta|X,Y) = 0.
\end{equation}
}
\end{itemize}

Actually it is the standard adiabatic procedure which has to lead to
the Schr\"{o}dinger equation for the slow electron motion
\begin{equation}\label{letsred}
  i\hbar\frac{\p}{\p t}\Phi(X) = H\Phi(X),
\end{equation}
with the effective slow motion Hamiltonian
\begin{equation}\label{hamslow}
  H = V(X,Y) + E(X,Y).
\end{equation}

However, there are some peculiarities caused by the fact that according to
Eqs.~(\ref{comrulefast},\ref{komsall}) neither fast nor slow variables
commute each with other. That is why both wave function parts in Eq.~(\ref{bangfunk})
depend only on a single variable (either $\eta$ or $X$), while the other one
has to be treated as an operator ($\xi=-i\p/\p\eta$, $Y=-l_B^2\p/\p X$).
Consequently, $X$ and $Y$ variables entering the fast wave function part
$\psi(\eta|X,Y)$ and the corresponding eigenvalue $E(X,Y)$ can not be treated
as parameters (what is done in the standard adiabatic procedure), but should
be considered as the operators acting on the slow wave function part.
This makes the adiabatic procedure a little bit tricky and cumbersome.
Nevertheless due to the presence of the small parameter $l_B/l_0$ it can
be performed. The details of this derivation are presented in the
Appendix~\ref{smh}. Restricting the consideration up to the $l_B^2$ order
we shall use the following slow motion Hamiltonian:
\begin{equation}\label{hamilton}
  H = V^{(S)}(\bi{R}) + \frac{l_B^2}{4}\nabla^2V^{(S)}(\bi{R}).
\end{equation}
The superscript $^{(S)}$ indicates that the expression should be
symmetrized in respect of the permutation of the slow variables $X$
and $Y$ which as we know already do not commute each with other.

The above adiabatic procedure can be easily generalized for the case
of many electron system. As the slow motion different electron
coordinates $\bi{R}_i$ commute each with other this generalization reduces
to inserting the proper summations into obtained slow motion Hamiltonian
and replacing it by the following expression:
\begin{equation}\label{hamiltonN}
  H = V^{(S)}(\bi{R}_1,\bi{R}_2,\cdots)
  + \frac{l_B^2}{4}\sum_{i=1}^N\nabla_i^2V^{(S)}(\bi{R}_1,\bi{R}_2,\cdots).
\end{equation}

Now we are going to consider some simple examples in order to
illustrate the application of the proposed simplified description
of the motion of electrons in the case of strong magnetic fields.
Let us start with the zero order ($l_B=0$) approximation.

\section{Classical equations of motion}

In zero order approximation we shall take into account only the
first term in Hamiltonian (\ref{hamilton}) and neglect the commutator
(\ref{komsall}) between $X$ and $Y$ coordinates.
We know that neglecting the commutators we have to arrive to
the classical mechanics. But one has to remember that it is not correct just
to neglect the commutators. It is necessary to replace them by the corresponding
Poisson brackets according to the following rule (note we inserted
$l_B^2$ instead of $\hbar$):
\begin{equation}\label{poisson}
  \frac{i}{l_B^2}[A,B] \quad \to \quad \{A,B\} =
  \frac{\p A}{\p y}\frac{\p B}{\p x}
  - \frac{\p A}{\p x}\frac{\p B}{\p y}.
\end{equation}
The most simple way to obtain the classical equations of motion is to use the
Heisenberg equations of motion for the operators. Thus, we write
\begin{eqnarray}
  \frac{d}{dt}X &=& \frac{i}{\hbar}[H,X]
  = \frac{l_B^2}{\hbar}\frac{i}{l_B^2}[H,X] \nonumber \\
  &\to& \frac{l_B^2}{\hbar}\{H,X\} = \frac{c}{eB}\frac{\p V}{\p Y}, \\
  \frac{d}{d t}Y &=& -\frac{c}{eB}\frac{\p V}{\p X}.
\end{eqnarray}
Note in the Heisenberg equations of motion the Plank constant $\hbar$ is used
(in spite of the fact that commutator of the variables is proportional
to the magnetic length squared), because it has to be in agreement with
the slow motion Schr\"{o}dinger equation (\ref{letsred}).
Those two equations of motion can be rewritten as a single vector equation
\begin{equation}\label{vectlg}
  \dot{\bi{R}} = -\frac{c}{eB}[\bi{e}_z\times\nabla]V(\bi{R})
\end{equation}
where the symbol $\bi{e}_z$ stands for the unit vector perpendicular
to the electron motion plane $z=0$.
It is well known equation in plasma physics, and it describes the
Larmor circle (the rotating electron in a strong magnetic field)
drift caused by the gradient of applied additional potential.

Thus, we see that system of 2D electrons in the very strong magnetic
field (in the conditions of the fractal Hall effect, when only the part
of the lowest Landau level is populated) demonstrates the classical
behavior. This classical behavior is rather tricky. They do not
behave as electrons. They behave as a system of classical gyroscopes.

Now let us go back and take the $l_B^2$ order terms into account.
In this case the quantum mechanical correction should take place,
and we have to obtain something like quasi-classical description.
In order to understand the main features of such quasi-classical
motion let us take the most simple example of the parabolic dot
with one and two electrons.

\section{Single electron in a parabolic dot}
\label{single}

In order to check the correctness of the above described method
let us start with trivial problem of a single electron in a parabolic
dot. In this case we have the following potential:
\begin{equation}\label{singlepot}
  V(\bi{r}) = \frac{m\omega_0^2}{2}r^2
\end{equation}
with the frequency $\omega_0$ characterizing the strength of the
confining potential, and according to Eq.~(\ref{hamilton})
the following slow motion Hamiltonian:
\begin{eqnarray}\label{seham}
  H &=& \frac{m\omega_0^2}{2}(X^2+Y^2) + \frac{1}{2}m\omega_0^2l_B^2 \nonumber \\
  &=& \frac{m\omega_0^2}{2}\left\{-l_B^4\frac{\p^2}{\p X^2}+X^2
  + \frac{1}{2}l_B^2\right\}.
\end{eqnarray}
This the well known Hamiltonian of the harmonic oscillator. Its
eigenvalues and the corresponding eigenfunctions are
\begin{eqnarray}\label{seewf}
\label{seewf1}
  E_n &=& \frac{\hbar\omega_0}{\gamma}(n+1), \\
\label{seewf2}
  \Phi_n(X) &=& \frac{1}{\sqrt{l_B2^nn!\sqrt{\pi}}}\e^{-X^2/2l_B^2}H_n(X/l_B)
\end{eqnarray}
where the parameter $\gamma=\omega_c/\omega_0$ characterizes the relative
strength of the magnetic field, and the symbol $H_n$ stands for the
Hermit polynomial.

\begin{figure}
\begin{center}
\setlength{\unitlength}{1cm}
\begin{picture}(8,7)
\put(0,0){\epsfig{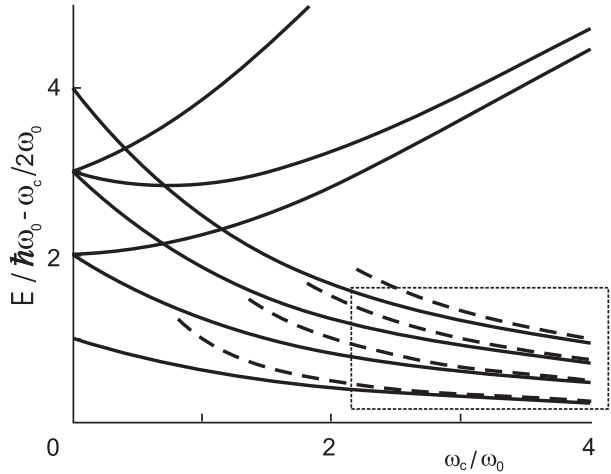}}
\end{picture}
\end{center}
\caption{Electron spectrum in a parabolic dot: solid curves --- the
exact result according (\ref{seexact}), dashed curves ---
the slow motion approximation (\ref{seewf1}).}
\label{fig1}
\end{figure}

In order to evaluate the approximation obtained by solving the slow
motion Schr\"{o}dinger equation let us compare it with the exact
Fock-Darvin result which is
\begin{equation}\label{seexact}
  E_{nm} = \hbar\omega_0\left\{(2n+|m|+1)\sqrt{1+\gamma^2/4}+(m-1)\gamma/2\right\}
\end{equation}
where orbital quantum number $m$ is integer, and radial quantum number $n$
is integer and nonnegative. This exact result together with the approximate
one (\ref{seewf1}) are shown in Fig.~\ref{fig1} by solid and dashed curves,
correspondingly. We see that in the asymptotic region $\gamma\to\infty$
(shown by the dotted rectangular) the approximate result is rather close to the
rotational levels belonging to the lowest Landau level. Moreover, we may expect
the quantitative agreement already at $\gamma\gtrsim 2$ values.
It is interesting to inspect how the wave function and the corresponding
electron density looks like. However, we have to remember that eigenfunction
(\ref{seewf2}) is not the electron wave function itself but it is its slow motion
part only. In order to obtain the electron wave function according to
Eq.~(\ref{bangfunk}) we have to multiply it by fast motion part.
Next we have to go back to the initial variables
(\ref{oldvar}). It can be done using some integral transformation which
is described in Appendix~\ref{vartransf}. Using transformation kernel
(\ref{tm}) and restricting our consideration by the lowest fast wave function
approximation (\ref{zeroorderenergy}) we write down the total electron
wave function in initial $x,y$ variables
\begin{eqnarray}\label{wftot}
&&  \Psi_n(x,y) = \int_{-\infty}^{\infty}d\eta\int_{-\infty}^{\infty}dX\,
  \langle x,y|\eta,X\rangle\psi_0(\eta)\Phi_n(X) \nonumber \\
&&  = \frac{1}{2\pi l_B\sqrt{2^nn!}}\int_{-\infty}^{\infty}d\eta
  \int_{-\infty}^{\infty}dX\,H_n(X/l_B) \nonumber \\
&&  \phantom{m}\cdot\e^{-\eta^2/2+iy(X-l_B\eta)/2l_B^2-X^2/2l_B^2}\,
  \delta(X+l_B\eta-x) \nonumber \\
&&  = \frac{\e^{(ixy-x^2)/2l_b^2}}{2\pi l_B\sqrt{2^nn!}}
  \int_{-\infty}^{\infty}d\eta\,
  \e^{-\eta^2+(x-iy)\eta/l_B}H_n(x/l_B-\eta). \nonumber \\
\end{eqnarray}
Fortunately the integral can be calculated by analytical means. Using
the standard integrals with Hermit polynomials \cite{ryzik} we obtain
the following expression for the total electron wave function
\begin{equation}\label{wftotfin}
  \Psi_n(x,y) = \frac{1}{l_B\sqrt{2^{n+1}n!\pi}}\e^{in\varphi}
  (r/l_B)^{n}\e^{r^2/4l_B^2},
\end{equation}
and the corresponding electron density in the $n$ eigenstate
\begin{equation}\label{density}
  \rho_n(\bi{r}) \sim (r/l_B)^{2n}\e^{-r^2/2l_B^2}.
\end{equation}

We see that in the case of large $n$ values (in the quasi-classical case)
the electrons are mainly located on the ring. Equating to zero
the derivative of the above density expression we obtain the radius
of this ring. It is
\begin{equation}\label{radiusring}
  r_0 = l_B\sqrt{2n}.
\end{equation}
Now inserting the $n$ value expressed from Eq.~(\ref{seewf1}) we get
\begin{equation}\label{radiusenergy}
  r_0 = l_B\sqrt{\frac{2\gamma E}{\hbar\omega_0}}
  = \sqrt{\frac{2E}{m\omega_0^2}},
\end{equation}
what exactly corresponds to the classical potential energy
$E=V(r_0)=m\omega_0^2r_0^2/2$ of the rotating electron drifting in the
confining potential along the circle with the radius $r_0$.
The single difference of quasi-classical electron behavior from the
classical one is that now according to Eq.~(\ref{density}) it moves
not along the thin trajectory, but it is spread over the ring with
the thickness of order $l_B$.

\section{Two electrons in a dot}

The other good for us example is the two electrons in a parabolic
dot because there is the exact solution which can be compared with our
approximate results (see, for instance, \cite{merkt91}).
In this case the behavior of electrons is described by the
following potential:
\begin{equation}\label{pottwo}
  V(\bi{r}_1,\bi{r}_2) = \frac{m\omega_0^2}{2}\{r_1^2+r_2^2\}
  + \frac{e^2}{|\bi{r}_1-\bi{r}_2|}.
\end{equation}
Let us introduce the center of mass and relative motion coordinates.
We shall do it in a non standard way in order not to spoil the commutations
rules for the fast and slow variables which we already used. Namely,
we use the following definition:
\begin{equation}\label{cmrelvar}
  \bi{r}_{\mathrm{c}} = \frac{1}{\sqrt{2}}(\bi{r}_1 + \bi{r}_2), \quad
  \bi{r}_{\mathrm{r}} = \frac{1}{\sqrt{2}}(\bi{r}_1 - \bi{r}_2).
\end{equation}
It leads to the separation of variables as the potential can be
presented as a sum of two terms
\begin{eqnarray}\label{pottwosplit}
  V(\bi{r}_1,\bi{r}_2) = V_{\mathrm{c}}(\bi{r}_{\mathrm{c}}) +
  V_{\mathrm{r}}(\bi{r}_{\mathrm{r}}).
\end{eqnarray}
The potential for the center of mass motion
\begin{equation}\label{cmpot}
  V_{\mathrm{c}}(\bi{r}_{\mathrm{c}}) = \frac{m\omega_0^2}{2}r_{\mathrm{c}}^2
\end{equation}
exactly coincides with the single electron potential (\ref{singlepot})
which was already considered in the previous Section.
Consequently, the eigenvalue and eigenfunction of the center of mass
motion coincide with those given by Eqs.~(\ref{seewf1},\ref{seewf2}).
Note that now the capitals $X$ and $Y$ have to be replaced by
the slow center of mass motion coordinates $X_{\mathrm{c}}$ and $Y_{\mathrm{c}}$.
Performing the same procedure as in the previous Section we shall
arrive at center of mass motion density given by Eq.~(\ref{density})
with the coordinate $\bi{r}$ replaced by the center of mass coordinate
$\bi{r}_{\mathrm{c}}$.

The relative motion potential is given as follows:
\begin{equation}\label{relpot}
  V_{\mathrm{r}}(\bi{r}_{\mathrm{r}})
  = \frac{m\omega_0^2}{2}r_{\mathrm{r}}^2 + \frac{e^2}{r_{\mathrm{r}}\sqrt{2}}.
\end{equation}
According to Eq.~(\ref{hamilton}) it leads to the following slow relative
motion Hamiltonian:
\begin{equation}\label{slowham}
  H_{\mathrm{r}} = \frac{m\omega_0^2}{2}\left(R_{\mathrm{r}}^2 + \frac{l_B^2}{2}\right)
  + \frac{e^2}{R_{\mathrm{r}}\sqrt{2}}\left(1+\frac{l_B^2}{4R_{\mathrm{r}}^2}\right).
\end{equation}
The symbol $R_{\mathrm{r}}^2$ of course has to be replaced by the operator
\begin{equation}\label{relrepl}
  R_{\mathrm{r}}^2 \to -l_B^4\frac{\p^2}{\p X_{\mathrm{r}}^2}+X_{\mathrm{r}}^2.
\end{equation}
Consistently the slow motion Schr\"{o}dinger equation with
Hamiltonian (\ref{slowham}) can be solved by means of Fourier transformation
technique presented in Appendix~\ref{smh}. But in this simple case of two electrons
one can find the eigenvalues of the above slow motion Hamiltonian
rather easily paying attention to the fact that the eigenfunctions
of $R_{\mathrm{r}}^2$ operator diagonalize Hamiltonian (\ref{slowham}) as well.
We know the eigenvalues and eigenfunctions of operator $R_{\mathrm{r}}^2$
already. They are given by Eqs.~(\ref{seewf1},\ref{seewf2}).
Thus, in order to obtain the eigenvalues of Hamiltonian (\ref{slowham})
we have to make the following replacement just in the above Hamiltonian:
\begin{equation}\label{eigenrep}
  R_{\mathrm{r}}^2 \to l_B^2(2n+1).
\end{equation}
Consequently, the relative slow motion eigenvalue reads
\begin{equation}\label{eigenvsr}
  E_n^{(r)} = \hbar\omega_0\left\{\frac{n+1}{\gamma}
  + \frac{\lambda\sqrt{\gamma}}{\sqrt{2(2n+1)}}\left[1+\frac{1}{4(2n+1)}
  \right] \right\}
\end{equation}
where dimensionless parameter of electron-electron interaction
$\lambda=l_0/a_B$ is the ratio of the characteristic confining
potential length $l_0=\sqrt{\hbar/m\omega_0}$ and the Bohr radius
$a_B=\hbar^2/me^2$.

\begin{figure}
\begin{center}
\setlength{\unitlength}{1cm}
\begin{picture}(8,6)
\put(0,0){\epsfig{file=fig2a.eps,width=6.8cm}}
\end{picture}
\end{center}
\begin{center}
\setlength{\unitlength}{1cm}
\begin{picture}(8,7)
\put(0,0){\epsfig{file=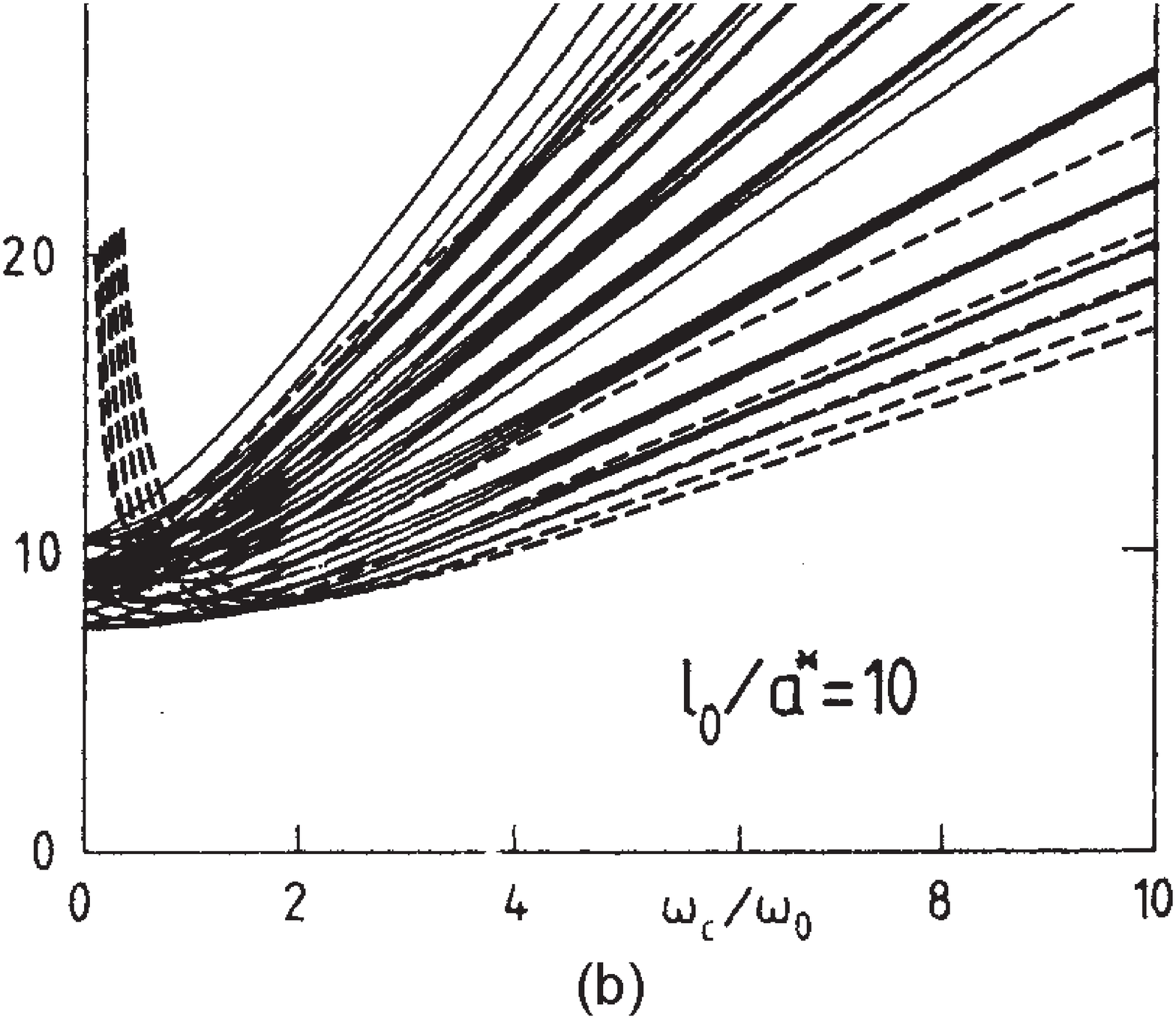,width=7cm}}
\end{picture}
\end{center}
\caption{Spectrum of two electron system in a parabolic dot:
a --- slow motion approximation results; b --- the comparison
of approximate results (dashed curves) with the exact results
(solid curves) taken from \cite{merkt91}.}
\label{fig2}
\end{figure}

Now adding together eigenvalue (\ref{seewf1}) for center of mass motion,
the relative motion eigenvalue (\ref{eigenvsr}), and one more term
$\hbar\omega_c$ for lowest Landau level energy we obtain the final
result for the two electron eigenvalue in the parabolic dot in the
slow motion approximation
\begin{eqnarray}\label{twosm}
  E_{N,n} &=& \hbar\omega_0\Bigg\{\gamma + \frac{N+n+2}{\gamma} \nonumber \\
&&  + \frac{\lambda\sqrt{\gamma}}{\sqrt{2(2n+1)}}\left[1+\frac{1}{4(2n+1)}
  \right] \Bigg\}.
\end{eqnarray}
The dimensionless eigenvalue (in $\hbar\omega_0$ units) dependencies
on the relative magnetic field strengths (on parameter $\gamma=\omega_c/\omega_0$)
for the case of $N=0$ and several $n$ values are shown in Fig.~\ref{fig2}a.
%
%\begin{widetext}
%
%
%\end{widetext}
%
In Fig.~\ref{fig2}b these eigenvalues are compared with the exact
solution taken from Merkt paper \cite{merkt91}.
We see that when $n\to\infty$ and $\gamma\to\infty$ (namely, in the
asymptotic region) the approximate consideration is in good agreement
with the exact one. Moreover, the quasi-classical treatment describes
correctly the main features of the electron behavior in strong
magnetic fields, namely, the increment of the angular momentum
(the quantum number $n$ plays its role) with
the increment of the magnetic field strength.
Indeed, minimizing relative motion eigenvalue (\ref{eigenvsr})
in respect the magnetic field $\gamma$ in the case of large
quantum numbers $n$ we obtain the ground state orbital momentum
\begin{equation}\label{orbital}
  n_0 = (\lambda/4)^{2/3}\gamma
\end{equation}
proportional to the magnetic field what agrees with the quantum mechanical
result obtained in \cite{matulis01}.

Let us also look at the electron density given by the quasi-classical
approximation. First, we notice that according what was said above
the eigenfunction of Hamiltonian (\ref{slowham}) coincides with the
eigenfunction of the operator $R_{\mathrm{r}}^2$. Thus, it coincides with
wave function (\ref{seewf2}) with the coordinate $X$ replaced by the
relative motion coordinate $X_{\mathrm{r}}$. So,
performing the same transformation
as it was done in Section \ref{single} we obtain the relative motion
density given by expression (\ref{density}).

\begin{figure}
\begin{center}
\setlength{\unitlength}{1cm}
\begin{picture}(8,6)
\put(0,0){\epsfig{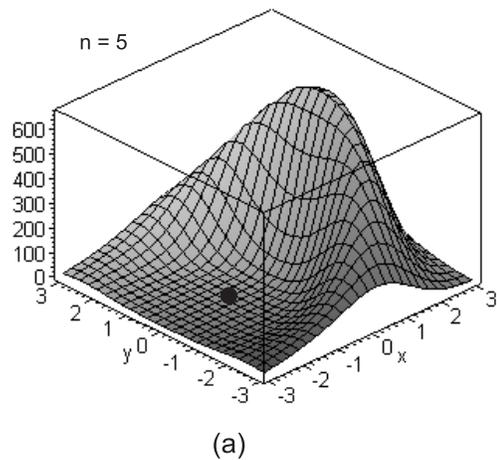}}
\end{picture}
\end{center}
\begin{center}
\setlength{\unitlength}{1cm}
\begin{picture}(8,7)
\put(0,0){\epsfig{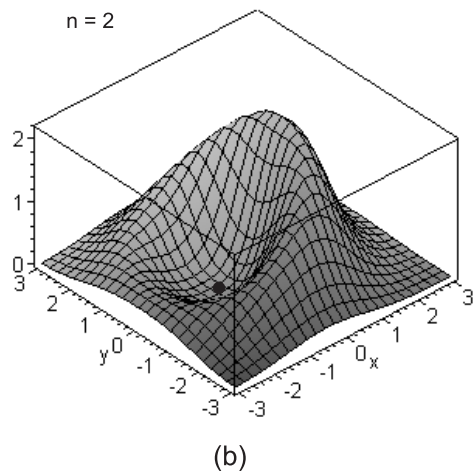}}
\end{picture}
\end{center}
\caption{Pair-correlation function for various orbital momenta.}
\label{fig3}
\end{figure}

Now taking into account that $N=0$ corresponds to the ground two electon
state we can write down the two electron distribution function in
the following form:
\begin{eqnarray}\label{twodistr}
  \rho_n(\bi{r}_1,\bi{r}_2) &\sim& \e^{-r_{\mathrm{c}}^2/2l_B^2}\cdot
  (r_{\mathrm{r}}/l_B)^{2n}\e^{-r_{\mathrm{r}}^2/2l_B^2} \nonumber \\
  &\sim& (\bi{r}_1-\bi{r}_2)^{2n}\e^{-(r_1^2+r_2^2)/2l_B^2}.
\end{eqnarray}
In Fig.~\ref{fig3} the above function is plotted as a function of the
first electron coordinate with the other electron coordinate fixed at the
point corresponding to its classical equilibrium position. This point
is indicated by solid dot.
Actually the plot represents so called pair correlation function.
The plot in Fig.~\ref{fig3}a corresponds to $n=5$ and the plot
in Fig.~\ref{fig3}b --- to $n=2$. We see that for larger $n$ the
pair correlation function demonstrates the peak in the opposite
to the fixed electron position what corresponds to the Wigner
crystallization of this simple two electron system
in the strong magnetic field.

When the magnetic field strength decreases (what corresponds to
the ground state with the smaller angular momentum value, say, $n=2$
as it is shown in Fig.~\ref{fig3}b) the Wigner crystal starts to melt
--- the pair correlation function transforms itself from the peak
into ring. Note, it demonstrates the fact that the angular melting
precedes the radial one what agrees with quantum mechanical result
in \cite{filinov00} obtained for the case without the magnetic field.

\section{Conclusions}

Let us summarize shortly what was said about the behavior of
electrons in the case of the strong magnetic field.

In the asymptotic region of the strong magnetic fields, when all
electrons occupy the lowest Landau level only, their kinetic energy
is frozen out, and their behavior is guided by weakly
varying (characterized by some characteristic length $l_0$)
additional potential. Applying special fast and slow motion variables,
the adiabatic procedure and the expansion in $l_B/l_0$
powers some simplified approximate description can be developed.

In zero order approximation one get the classical equations which
describe the electrons as a system of gyroscopes. Those equations
actually are the equations for the Larmor circle drift in the
gradient of applied potential.

Applying the expansion up to $(l_B/l_0)^2$ order one obtain
the self consistent equation set which coincides with the Schr\"{o}dinger
equation where the role of canonical variables play two cartesian slow
motion 2D electron coordinates $X$ and $Y$ with the commutator proportional
to the magnetic length squared (instead of being proportional to the
Plank constant as it is in the standard Schr\"{o}dinger equation).

Two simple examples of a single and two electron in a parabolic dot
demonstrate the accuracy and main features of proposed approximate
description.

The approximate eigenvalues coincide with the exact ones in the asymptotic
region when $\gamma=\omega_c/\omega_0\to\infty$, and even in the intermediate
region ($\gamma\gtrsim 2$) one can expect rather good semi-quantitative
description.

The electron wave functions can be obtained as product of the fast wave
function part corresponding to the lowest Landau level and slow wave
function part obtained by the above specific slow motion Schr\"{o}dinger
equation. After the transformation to the original variables the
wave function obtained in this way describes correctly the quantum
mechanical $l_B/l_0$ order correction to the classical electron
motion.

The two electron in a dot example shows that the proposed approximate
consideration describes such collective phenomena as the Wigner
crystallization, the change of the angular momentum in the ground
state when the magnetic field strength increases, the phenomena of
angular and radial melting of the Wigner crystal. We hope that this
approximate method can be useful for the consideration of more
sophisticated many electron systems
when the straightforward solution of quantum mechanical equation meets
computational difficulties.

\begin{acknowledgments}
I would like to acknowledge Prof.~Fran\c{c}ois Peeters and
Dr.~Bart Partoens from the Antwerp University. Most my ideas on
quantum dots appeared during my numerous visits there, due to the close
collaboration and the discussions with them. I would like to thank
Dr.~Egidijus Anisimovas for drawing my attention to various representations
of electron wave functions in magnetic field.
\end{acknowledgments}

\appendix

\section{Slow motion Hamiltonian}
\label{smh}

As it was mentioned in Section \ref{adiabatproc} performing the
adiabatic procedure steps we have to pay attention to the fact that
the variables $X$ and $Y$ do not commute each with other. That is why
instead of using the straightforward expansion (\ref{letpotskl})
we apply the following Fourier transformation:
\begin{eqnarray}\label{furjetrans}
\label{furjetrans1}
  V(x,y) &=& \frac{1}{2}\int_{-\infty}^{\infty}\frac{dk}{2\pi}
  \int_{-\infty}^{\infty}\frac{dq}{2\pi} \nonumber \\
&&  \phantom{m}\cdot\Big\{\e^{ikx}\e^{iqy}
  + \e^{iqy}\e^{ikx}\Big\}V(k,q), \\
\label{furjetrans2}
  V(k,q) &=& \int_{-\infty}^{\infty}dx \int_{-\infty}^{\infty}dy
  \e^{-ikx}\e^{-iqy}V(x,y).
\end{eqnarray}
These two expressions can be considered as a definition of the
operator function $V(x,y)$. Thus in the first expression the symbols
$x$ and $y$ will be considered as the operators, while in the second
one $x$ and $y$ are just the dummy integration variables.
The main advantage of such potential representation is that the
operators $x$ and $y$ are moved from the general potential function
$V(\bi{r})$ to more simple exponent functions. Although the old $x$
and $y$ variables commute we used the symmetric exponent product
which will be necessary in further derivation.

Now we substitute variables (\ref{oldvar}) into exponents and expand
them into $\xi,\eta$-powers:

\begin{widetext}
\begin{eqnarray}
  \e^{ik(X+l_B\eta)}\e^{iq(Y-l_B\xi)}
  &=& \e^{ikX}\e^{iqY}\e^{il_Bk\eta}\e^{-il_Bq\xi} \nonumber \\
  &=& \e^{ikX}\e^{iqY} \Big\{1+il_Bk\eta-\frac{1}{2}l_B^2k^2\eta^2\Big\}
  \Big\{1-il_Bq\xi-\frac{1}{2}l_B^2q^2\xi^2\Big\} \nonumber \\
  &=& \e^{ikX}\e^{iqY} \Big\{1+il_B(k\eta-q\xi)-\frac{1}{2}l_B^2k^2\eta^2
  -\frac{1}{2}l_B^2q^2\xi^2 + l_B^2kq\eta\xi\Big\}, \\
  \e^{iq(Y-l_B\xi)}\e^{ik(X+l_B\eta)}
  &=& \e^{iqY}\e^{ikX}\e^{-il_Bq\xi}\e^{il_Bk\eta} \nonumber \\
  &=& \e^{iqY}\e^{ikX} \Big\{1+il_B(k\eta-q\xi)-\frac{1}{2}l_B^2k^2\eta^2
  -\frac{1}{2}l_B^2q^2\xi^2 + l_B^2kq\xi\eta\Big\}.
\end{eqnarray}

Taking the evident equality
\begin{eqnarray}\label{aklyg}
  a\eta\xi + b\xi\eta = \frac{1}{2}a(\eta\xi+\xi\eta+i)
  + \frac{1}{2}b(\xi\eta+\eta\xi-i) %\nonumber \\
  = \frac{1}{2}(a+b)(\xi\eta+\eta\xi) + \frac{i}{2}(a-b)
\end{eqnarray}
into account we write down the following expansion of the symmetric
product of exponents:
\begin{eqnarray}
&& \e^{ik(X+l_B\eta)}\e^{iq(Y-l_B\xi)}
  + \e^{iq(Y-l_B\xi)}\e^{ik(X-l_B\eta)} %\nonumber \\
%&&
  = \Big\{\e^{ikX}\e^{iqY}
  + \e^{iqY}\e^{ikX}\Big\} \nonumber \\
&& \phantom{mmmm} \times\Big\{1+il_B(k\eta-q\xi)
  -\frac{1}{2}l_B^2(k\eta-q\xi)^2\Big\} %\nonumber \\
%&& \phantom{mm}
  + \frac{i}{2}l_B^2kq\Big\{\e^{ikX}\e^{iqY}
- \e^{iqY}\e^{ikX}\Big\} \nonumber \\
&&  = 2\Big[\e^{ikX}\e^{iqY}\Big]^{(S)} L(\xi,\eta,k,q)
  +il_B^2kq\Big[\e^{ikX}\e^{iqY}\Big]^{(A)}. \nonumber \\
\end{eqnarray}
Note how the symmetric and antisymmetric exponent products and
function $L(\xi,\eta,k,q)$ are defined.

Inserting the above expansion into Fourier transformation (\ref{furjetrans}),
changing the parameters $k$ and $q$ by the operators $i\p/\p x$ and $i\p/\p y$
acting on exponents, and performing the integration by parts we arrive
at the following potential expansion:
\begin{eqnarray}\label{potskl}
&&  V(X+l_B\eta,Y-l_B\xi) = \int_{-\infty}^{\infty}dx \int_{-\infty}^{\infty}dy
  \int_{-\infty}^{\infty}\frac{dk}{2\pi}
  \int_{-\infty}^{\infty}\frac{dq}{2\pi}V(x,y) \nonumber \\
&&
  \phantom{mmmmmm} \times\Bigg\{L(\xi,\eta,k,q)
  \Big[\e^{ik(X-x)}\e^{iq(Y-y)}\Big]^{(S)}
  +\frac{i}{2}l_B^2kq\Big[\e^{ik(X-x)}\e^{iq(Y-y)}\Big]^{(A)}\Bigg\} \nonumber \\
&&
  = \int_{-\infty}^{\infty}dx \int_{-\infty}^{\infty}dy
  \int_{-\infty}^{\infty}\frac{dk}{2\pi}
  \int_{-\infty}^{\infty}\frac{dq}{2\pi}V(x,y) %\nonumber \\
%&&
%  \phantom{m}
  \times\Bigg\{L(\xi,\eta,i\p/\p x,i\p/\p y)
  \Big[\e^{ik(X-x)}\e^{iq(Y-y)}\Big]^{(S)} \nonumber \\
&& \phantom{mmmmmm} -\frac{il_B^2}{2}\frac{\p^2}{\p x\p y}
  \Big[\e^{ik(X-x)}\e^{iq(Y-y)}\Big]^{(A)}\Bigg\} \nonumber \\
&&  = \int dx \int dy   \int\frac{dk}{2\pi}\int\frac{dq}{2\pi}
  \Bigg\{\Big[\e^{ik(X-x)}\e^{iq(Y-y)}\Big]^{(S)}
  L(\xi,\eta,-i\p/\p x,-i\p/\p y) \nonumber \\
&&
  \phantom{mmmmmm}
  -\frac{il_B^2}{2}\Big[\e^{ik(X-x)}\e^{iq(Y-y)}\Big]^{(A)}
  \frac{\p^2}{\p x\p y}\Bigg\}V(x,y) \nonumber \\
&&  = \int dx \int dy  \int\frac{dk}{2\pi}\int\frac{dq}{2\pi}
  \Bigg\{\Big[\e^{ik(X-x)}\e^{iq(Y-y)}\Big]^{(S)}\Big\{1+
  l_B\Big(\eta\frac{\p}{\p x}-\xi\frac{\p}{\p y}\Big) %\nonumber \\
  - \frac{l_B^2}{2}\Big(\eta\frac{\p}{\p x}
  -\xi\frac{\p}{\p y}\Big)^2\Big\} \nonumber \\
&& \phantom{mmmmmm}  -\frac{il_B^2}{2}\Big[\e^{ik(X-x)}\e^{iq(Y-y)}\Big]^{(A)}
  \frac{\p^2}{\p x\p y}\Bigg\}V(x,y).
\end{eqnarray}
Actually it is the definition of the operator function expansion which
we have to use instead of expression (\ref{letpotskl}). It can be rewritten
in more simple way if we use the following operator function definition:
\begin{equation}\label{duop}
  F^{(XY)}(X,Y) = \int dx\int dy\int\frac{dk}{2\pi}\int\frac{dq}{2\pi}
  \e^{ik(X-x)}\e^{iq(Y-y)}F(x,y).
\end{equation}
\end{widetext}
It defines the function with ordered operators --- all operators $X$ stand
on the left side of the operators $Y$ in all terms of its Taylor expansion,
or Fourier transform.

Defining the symmetric and antisymmetric operator function as
\begin{eqnarray}\label{operfunk}
  F^{(S)}(X,Y) = \frac{1}{2}\Big\{F^{(XY)}(X,Y)+F^{(YX)}(X,Y)\Big\}, \\
  F^{(A)}(X,Y) = \frac{1}{2}\Big\{F^{(XY)}(X,Y)-F^{(YX)}(X,Y)\Big\}
  \phantom{n}
\end{eqnarray}
we rewrite the potential expansion in the following formal simple form:
\begin{eqnarray}\label{potkomp}
  V &=& \Bigg\{1 + l_B\Big(\eta\frac{\p}{\p X}
  -\xi\frac{\p}{\p Y}\Big) \nonumber \\
  &-& \frac{l_B^2}{2}\Big(\eta\frac{\p}{\p X}
  -\xi\frac{\p}{\p Y}\Big)^2\Bigg\}V^{(S)}
%  \nonumber \\
  + \frac{il_B^2}{2}V_{XY}^{(A)}.\phantom{m}
\end{eqnarray}

Now we are ready to perform the next step of our adiabatic procedure,
namely, to insert the obtained potential expansion into fast Hamiltonian
(\ref{hampad2}) and solve the fast eigenvalue problem (\ref{greitlg}).
It can be easily performed using the standard perturbation technique.

Indeed, using the modified fast Hamiltonian
\begin{eqnarray}\label{fastham}
  H_f &=&  H_0 + H_1 + H_2, \\
  H_0 &=& \frac{\hbar\omega_c}{2}(\xi^2+\eta^2), \\
  H_1 &=& l_B\Big(\eta\frac{\p}{\p X}-\xi\frac{\p}{\p Y}\Big)V^{(S)}, \\
  H_2 &=& - \frac{l_B^2}{2}\Big(\eta\frac{\p}{\p X}
  -\xi\frac{\p}{\p Y}\Big)^2V^{(S)}
\end{eqnarray}
we obtain the zero order eigenvalue and function
\begin{equation}\label{zeroorderenergy}
  E_0 = \frac{\hbar\omega_c}{2}, \quad \psi_0(\eta)
  = \pi^{-1/4}\e^{-\eta^2/2}.
\end{equation}
Next, due to the zero order function symmetry we get $E_1=0$, and solving
the first order equation
\begin{eqnarray}\label{firstorder}
  \left\{H_0-E_0\right\} = -H_1\psi_0 =
  - l_B(V_X^{(S)}-iV_Y^{(S)})\eta\psi_0 \phantom{m}
\end{eqnarray}
we define the first order correction to the wave function
\begin{equation}\label{firstorderwf}
  \psi_1(\eta|X,Y) = - \frac{l_B(V_X^{(S)}-iV_B^{(S)})}{\hbar\omega_c}\eta\psi_0.
\end{equation}
Then from the second order equation one can easily
get the following second order eigenvalue correction:
\begin{eqnarray}\label{secorder}
&&  E_2(X,Y) = \int_{-\infty}^{\infty}d\eta\,
  \psi_0(\eta)H_2\psi_0(\eta) \nonumber \\
&& \phantom{mmmm}+ \int_{-\infty}^{\infty}d\eta\,
  \psi_0(\eta)H_1\psi_1(\eta|X,Y) \nonumber \\
&&  = \frac{l_B^2}{4}\left\{V_{XX}^{(S)}+V_{YY}^{(S)}\right\}
  - \frac{l_B^2}{2\hbar\omega_c}\left\{V_X^{(S)2}
  +V_Y^{(S)2}\right\}.\phantom{mmm}
\end{eqnarray}

Now having the fast motion problem eigenvalue calculated we can proceed
with the adiabatic procedure. For this purpose we present the total
wave function as
\begin{equation}\label{wfvisa}
  \Psi(\eta,X,t) = \e^{-iE_0t}\Big\{\psi_0(\eta) +
  \psi_1(\eta|X,Y)\Big\}\Phi(X,t),
\end{equation}
insert it into Eq.~(\ref{sred}), and obtain the following expression:
\begin{eqnarray}\label{atrlyg}
&&  \Big\{i\frac{\p}{\p t}+E_0-H_f-V^{(S)}-\frac{il_B^2}{2}V_{XY}^{(A)}\Big\} \nonumber \\
&& \phantom{m} \cdot \Big\{\psi_0(\eta) + \psi_1(\eta|X,Y)\Big\}\Phi(X,t) = 0.
\end{eqnarray}
Now multiplying the above equation by function $\psi_0(\eta)$ from the left
side, integrating it over all $-\infty<\eta<\infty$ interval and taking
into account the action of the fast Hamiltonian $H_f$ on the fast wave function
part we arrive at the slow motion equation (\ref{letsred}) with the
effective Hamiltonian
\begin{equation}\label{letham}
  H = V^{(S)} + \frac{il_B^2}{2}V_{XY}^{(A)}
  + \frac{l_B^2}{4}\nabla^2V^{(S)}
  -\frac{l_B^2}{2\hbar\omega_c}\{\nabla V^{(S)}\}^2.
\end{equation}
The performed procedure is consistent at least with the accuracy up to
$l_B^2$. That is why we shall omit the second and the last terms in
(\ref{letham}) as they are of order $l_B^4$. In the last term this
dependence appears due to the additional factor $\omega_c$ in the
denominator, while in the second term it is caused by slow variable
commutator (\ref{komsall}). Omitting these terms we arrive at the final
slow motion Hamiltonian (\ref{hamilton}).

\section{Coordinate transformation}
\label{vartransf}

According to the ideas of the quantum mechanics we can change the wave function
variables by means of the following transformation:
\begin{equation}\label{vartrans}
  \Psi(x,y) = \int_{-\infty}^{\infty}d\eta\int_{-\infty}^{\infty}dX\,
  \langle x,y|\eta,X\rangle \Psi(\eta,X)
\end{equation}
where the transformation function $\langle x,y|\eta,X\rangle$ has to be chosen
as the eigenfunction of operators $\hat{x}$ and $\hat{y}$ with the
corresponding eigenvalues $x$ and $y$. Namely, this transformation
function has to obey the following equations:
\begin{eqnarray}\label{treq}
  \{\hat{x}-x\}\langle x,y|\eta,X\rangle &=& 0, \\
  \{\hat{y}-y\}\langle x,y|\eta,X\rangle &=& 0,
\end{eqnarray}
or
\begin{eqnarray}\label{eqytransf}
  \left\{X+l_B\eta-x\right\}\langle x,y|\eta,X\rangle &=& 0, \\
  \left\{-il_B^2\frac{\p}{\p X}+il_B\frac{\p}{\p\eta}-y\right\}
  \langle x,y|\eta,X\rangle &=& 0.
\end{eqnarray}
It can be checked straightforwardly that the following transformation
function satisfies both equations:
\begin{equation}\label{tm}
  \langle x,y|\eta,X\rangle = \frac{1}{2\sqrt{\pi l_B}}
  \e^{iy(X-l_B\eta)/2l_B^2}\delta(X+l_B\eta-x).
\end{equation}
The normalization factor is chosen in agreement with the condition
\begin{eqnarray}\label{normfact}
&&  \int_{-\infty}^{\infty}dx\int_{-\infty}^{\infty}dy\,
  \langle x,y|\eta,X\rangle\langle x,y|\eta',X'\rangle \nonumber \\
&& \phantom{mm}  = \delta(\eta-\eta')\delta(X-X').
\end{eqnarray}

\end{document}